\documentstyle[12pt,aaspp4]{article}    
\newcommand{\be}{\begin{equation}}  
\newcommand{\ee}{\end{equation}}    
\newcommand{\bea}{\begin{eqnarray}}
\newcommand{\eea}{\end{eqnarray}}
\def\nAV{\langle n\rangle}
\def\GammaAV{\langle\Gamma\rangle}
\def\neff{n_{\rm e}}

\slugcomment{}
\lefthead{Kennedy \& Bludman}
\righthead{Kennedy \& Bludman}

\begin{document}

\begin{flushright}
UF-IFT-HEP-98-4\\ UPR-799T\\ NSF-ITP-98-125\\ 
Original: December 1998\\ Revised: June 1999
\end{flushright}
\title{Analytic Models for the Mechanical Structure of the Solar Core\footnote{
Based on a contribution to the Conference on Solar Neutrinos: News About SNUs, 
ITP/U. California at Santa Barbara, December 1997}}
\author{Dallas C. Kennedy}
\affil{Department of Physics, University of Florida, Gainesville FL 32611,\\
kennedy@phys.ufl.edu}
\and
\author{Sidney A. Bludman}
\affil{Department of Physics and Astronomy, University of Pennsylvania,
Philadelphia PA 19104\\ and}
\affil{DESY, Hamburg D-22607 Germany, bludman@mail.desy.de} 
\begin{abstract}
All stars exhibit universal central behavior in terms of new
homology variables $(u,w).$  In terms of these variables,
we obtain simple analytic fits to numerical standard solar models
for the core and radiative zones of the ZAMS and present 
Suns, with a few global parameters.  With these analytic fits, 
different theoretical models of the solar core, neutrino fluxes,
and helioseismic observations can be parametrized and compared.
\end{abstract}
\bigskip
\centerline{To appear in {\bf the Astrophysical Journal}: 10 November 1999}
\bigskip
\indent The complete static structure of a star of given composition profile
involves two equations of
mechanical equilibrium coupled to two equations of energy equilibrium.  These
must be solved simultaneously to high precision, in order to predict neutrino
fluxes and helioseismology.  In this paper, we do not solve for a standard
solar model (SSM) but instead fit a known  
standard solar model and helioseismic observations to
analytical formulae, so as to understand simply the 
the energy- and neutrino-producing solar core.

We are helped by the fact that in cool, lower
Main Sequence stars, temperature gradients are small.  This practically 
decouples the centrally-concentrated inner core from the star's 
extended radiative zone
and convective envelope.  Within the cores of such
centrally-concentrated stars, the density and pressure profiles are,
except for scale, determined by mass
conservation and hydrostatic equilibrium.  The other two equations, for
energy production and transport, are essential in determining the
thermal structure and the {\em normalization} of core mass and
radius or, equivalently, of central density and pressure $(\rho_c,P_c).$

We use homology invariants to isolate the dimensionless mechanical
structure of any star whose barytropic structure $P(\rho)$ is known.
We do {\em not} assume homology for
the Sun or any part of it, but use the regular (E) solutions of the
Lane-Emden equation only to illustrate our method.

Our fits do not describe the thermal structure or the whole Sun.  Indeed, we
must take the central density and pressure $(\rho_c,P_c)$ from complete
solar models or from helioseismic data.  
Because lower Main Sequence stars are nearly ideal gases, the ideal gas
law $P/\rho=\Re T/\mu$  determines $T/\mu$ or the core's thermal profile,
once the composition profile is specified.  Our best fits (Section~4)
reproduce numerical SSMs to about a percent and can be used for the analysis
of more-or-less model-independent helioseismic data.  Indeed, 
helioseismology and the asteroseismology of Sun-like stars will ultimately 
{\em calibrate} solar models and may rule out severely non-standard SMs as a
solution of the solar neutrino problem.  
The SM class homologous to the SSM is already ruled out
because of its tight correlation between neutrino fluxes and central 
temperature, density, and nuclear cross sections (Bludman \& Kennedy 1996,
Hata \& Langacker 1997).

\section{Core Structure of Centrally-Concentrated Stars}
\subsection{Mechanical Structure Described by Homology Invariants}

Homology-invariant variables were introduced by Chandrasekhar 1939 and
Schwarzschild 1958 and are discussed, for example, by Cox \& Giuli
1968 and Kippenhahn \& Weigert 1990.  In place of the usual variables 
\begin{equation}
u\equiv {d\ln m\over{d\ln r}}\quad ,\quad v\equiv -{d\ln (P/\rho)\over{d\ln r}}
\quad \mbox{or} \quad V\equiv  -{d\ln P \over{d\ln r}} =(\neff +1)v \quad,
\end{equation}
we choose 
\begin{equation}
u\equiv {d\ln m\over{d\ln r}}\quad ,\quad w\equiv -{d\ln\rho\over{d\ln r}}
\equiv\neff v = {V\over\Gamma}\quad ,
\end{equation}
\noindent in order to demonstrate universal density behavior~(15) in the 
inner core. Here $\Gamma\equiv d\ln P/d \ln \rho \equiv 1+1/\neff$ 
defines the local
polytropic index $\neff = d\ln \rho/d\ln (P/\rho)$, which is constant 
only for a polytrope.  Although stellar structure is far from being 
self-homologous (polytropic), these homology-invariant variables have 
nevertheless historically proven to be useful.

Because the Sun is well-approximated by an ideal gas equation of state,
$1-\Gamma^{-1}\equiv 1/(\neff +1)$ = $d\ln(T/\mu )/d\ln P$ = $\nabla 
-\nabla_{\mu}$,
where $\nabla\equiv d\ln T/d\ln P$ and
$\nabla_{\mu}\equiv d\ln {\mu}/d\ln P$ are the
temperature and mean molecular weight logarithmic gradients.
The barytrope $P(\rho)$ and $\neff$ we are fitting both derive from
the star's thermal structure and chemical composition.  For cool stars 
like the Sun, $\neff$ is not constant, but is large:
the core is centrally concentrated and surrounded by extended
radiative and convective zones.  This fixes, to good approximation,
the mechanical structure of the {\em inner core} of the Sun and other stars
with extended envelopes.

The two mechanical equations
\begin{equation}
{dm\over{dr}}=4 \pi r^2 \rho\quad ,\quad -{1\over\rho}\cdot {dP\over{dr}} 
= g \equiv {Gm(r)\over{r^2}}\quad ,
\end{equation}
for mass continuity and for pressure equilibrium, can be written
respectively as
\begin{eqnarray}
du/d\ln m & = & 3-w-u\quad ,\\
d\ln V/d\ln r & = & -1 + u + w/\neff\quad ,
\end{eqnarray}
or
\begin{equation}
d\ln r=d\ln u/(3-w-u) = d\ln V/[-1+u+w/\neff]\quad .
\end{equation}
The last equation (hydrostatic equilibrium) is a first-order equation
for homology variables
\be
{d\ln (\Gamma w)\over{d\ln u}}={-1+u+w(\Gamma-1) \over{3-u-w}}\quad ,
\ee
provided $\Gamma$ is a given function of $u$ or $w$, determined
by the prescribed thermal structure or barytrope $P=P(\rho)$.
For continuously varying $\neff$ or $\Gamma$, this is
an extension of composite polytropes of different, constant
$\neff = n$ (Chandrasekhar 1939).

In Section~4, we obtain $u,~w$ from the present and ZAMS standard solar 
models, including helium and metal diffusion, calculated by Bahcall, 
Pinsonneault and collaborators, and called the BP98 models.  These are
Bahcall, Basu, \& Pinsonneault 1998 and Bahcall, Basu, Pinsonneault, \& 
Christensen-Dalsgaard 1997 for the present Sun, and Pinsonneault 1998, for
the ZAMS Sun.  (See also Bahcall 1999.)
We use the first (mass continuity) 
equation only to obtain the radial and mass distributions, $r(u),~m(u)$:
\begin{equation}
{dm\over{m}}=u \cdot {dr\over{r}}={dz\over{w(z)-z}}\quad ,
\end{equation}
where $z\equiv 3 - u$ = $-d\ln\overline{\rho}/d \ln r$ and
$\overline{\rho}\equiv m(r)/(4\pi r^3/ 3)$ is the average density
interior to radius $r$.  Since $d\ln g/d\ln r$ = $u-2=1-z$, the
gravitational acceleration $g\equiv Gm/r^2$ is a maximum at $u$ = 2, and
$u,~w$ have inflection points as functions of $r,~m$.
We define the {\em inner core} boundary as this radius: 
$u(\xi_{ic})\equiv 2$ or $z(\xi_{ic})\equiv 1$.

For regular solutions, $w\rightarrow 5z/3$ at $z$ = 0.  For solutions of 
{\em finite} radius, $w(z)$ diverges at the stellar surface  
\begin{equation}
du/d\ln m\rightarrow -w\quad ,\quad 
uV^{\neff} \rightarrow\ {\rm constant}\quad ,\quad z \rightarrow 3.
\end{equation}
For solutions of {\em infinite} radius, $w$ remains finite as
$r\rightarrow$ $\infty$.  A centrally-concentrated core is insensitive
to the stellar surface. In Section~3, we therefore neglect the outer 
boundary condition by letting $w$ be finite at $z$ = 3, so that the core 
radius recedes to infinity.

\subsection{Polytropes as an Illustration}

For comparison with analytic fits, we often study the polytropic
case, $\neff = n$ = constant, in which equation~(7) reads
\begin{eqnarray}
{d\ln w\over{d\ln u}}={{-1+u+w/n}\over{3-w-u}}\quad \ {\rm or}\ \quad
{d\ln w\over{dz}} = {{2-z+w/n}\over{(3-z)(w-z)}}\quad .
\end{eqnarray}
Figure~1 shows the
characteristic curves in the $(u,w)$ plane for complete polytropes
$P=K\rho^{\Gamma}$, in which $\Gamma
\equiv 1+1/n$ is a constant fixed by the star's temperature 
gradient and $K = P_c/\rho^\Gamma_c$ is another constant determined by
the star's total entropy or luminosity.

For polytropes, $w$ = $n(-\xi\theta^\prime_n/\theta_n)$, where ${}^\prime$ =
$d/d\xi$, so that~(3) give
\begin{eqnarray}
(\xi^2\theta_n ')'+\xi^2\theta^n_n = 0\quad ,\\
\theta=(P/\rho)/(P/\rho)_c\quad, \quad \xi^2 \equiv r^2/\alpha^2 \quad ,\quad
\alpha^2 \equiv (n+1) R_0^2/6 \equiv(n+1)P_c/4\pi G \rho_c^2 \quad ,\nonumber
\end{eqnarray}
the Lane-Emden equation of index $n$.  We consider only the E-solutions, 
complete polytropes regular at the origin, for which $\theta_n$ = 1, 
$\theta^\prime_n$ = 0 at $\xi$ = 0.

In a polytrope, $\xi^2\sim 5z/n\sim 3v$, so that $\xi^2_{ic}\sim 5/n$.
Therefore $\xi^2_{ic}$ is insensitive to $n$, for $3 < n < 5$.  The
stellar radius, however, diverges for $n=5$.  The constants of 
the Lane-Emden functions (Chandrasekhar 1939) imply
$\xi_1\simeq 3(n+1)/(5-n) \gg \xi_{ic}$, for $3 < n < 5$.
Figure~1 shows that over this range in $n$, polytropes enjoy a common
core structure and differ only by their more-or-less extended
envelopes.

The two limiting cases of~(10) are the polytropes $n = 0$
(incompressible matter) and $n = 5$ (the most centrally concentrated
star of finite mass $M$). For incompressible matter, $m(r)=(4\pi/3)\rho r^3$,
the exact solution is
\begin{equation}
\theta_0=1-\xi^2/6\quad ,\quad P/P_c =\theta_0\quad, \quad \rho/\rho_c=1\quad .
\end{equation}
For the $n=5$ polytrope,
\begin{eqnarray}
\theta_5=[1+\xi^2/3]^{-1/2}\quad ,\quad \rho/\rho_c=\theta_5^5\quad ,\quad 
P/P_c=\theta_5^6\quad .
\end{eqnarray}
In the $(u, w)$ plane, we have 
\begin{eqnarray}
u = \left\{ \begin{array}{l}
             3           \quad \\
             3(1-w/5)    \quad \\
            \end{array}\right. \quad ,\quad
z = \left\{ \begin{array}{ll}
             0           &  \mbox{for $n$ = 0}\quad  \\
             3w/5        &  \mbox{for $n$ = 5}\quad  \\
            \end{array}\right. \quad .
\end{eqnarray}
These two limiting cases are the only ones for which $u$ is linear in $w$.
Examination of Figure~1 shows that all polytropes, and therefore all
barytropes of finite mass, fall between these two limits.

The rest of this paper stresses the dominant role of mass continuity
in the inner core of centrally-concentrated stars and uses equations~(4--8) 
to extract the radial and mass profiles of the core from a realistic
barytrope, which we take from the BP98 models or from
helioseismic data.  We solve only the first two (mechanical) 
equations of stellar core structure.  The remaining two
(thermal) equations of stellar structure, not considered here, determine
the barytrope and luminosity and neutrino production.

\subsection{Mass Continuity Determines Inner Core Structure}

From mass continuity (8)
\begin{equation}
\overline{\rho}(r)=\rho_c^{2/5} \rho^{3/5}(r)\quad,
\end{equation}
to first order in $r^2$.
For centrally-condensed stars
$(\neff > 3),$ this limiting behavior
$w=(5/3)z$
remains a good approximation over the inner core $u>2$, and leads to
\begin{equation}
u(m) \simeq 3 - (m/M_{ic})^{2/3}\quad ,\quad m/M_{ic} \simeq 
(3 - u)^{3/2} \quad ,
\end{equation}
where $M_{ic}$ is the inner core mass.

For the $n=5$ polytrope, equation~(16) holds
exactly everywhere, the total mass $M$ is finite and independent of
the radius $R$, which is infinite.  In this special case, the entire star
is strictly homologous to the core.  In centrally-concentrated non-polytropic
stars, this remains approximately true:
the dimensionless core profile is insensitive to 
the outer boundary condition, which only fixes the core scale.

Although not polytropic, all cool stars enjoy this common inner core 
profile: their low luminosity makes the temperature gradients small.  
Once $(\rho_c, P_c)$ are specified, the core's
structure is virtually determined by mass continuity and pressure 
equilibrium, rather than by their small energy generation and radiative
transfer.

Defined this way, the
inner core incorporates 97\% (84\%) of the ZAMS (present) Sun's
luminosity production.  We therefore fit over the {\em entire core} 
($u>1,~z<2$), constituting essentially all of the luminosity-producing 
region and about 65--70\% of the Sun's mass.  The parameters
of the entire core are summarized in Table~2 and shown by vertical 
arrows in Figures~2--3.  At the outer core radius,
$w\sim 10/3$ for the present Sun and $\sim$ 4 for the ZAMS Sun.

We use Schwarzschild dimensionless radius and mass defined in terms of
the total stellar mass $M$ and radius $R$ (Schwarzschild 1958).  
When equation~(8) is integrated,
$$
m(z)/M = (z/3)^{3/2}\cdot\exp\Bigg[\int^{z}_{3}dz\Bigg\lbrace{1\over
{w(z)-z}}-{3\over{2z}}\Bigg\rbrace\Bigg] 
\equiv q(u)\quad ,\eqno(17a)
$$
$$
r(z)/R = (z/3)^{1/2}\cdot\exp\Bigg[\int^{z}_{3}dz\Bigg\lbrace{1\over{(3-z)
[w(z)-z]}}-{1\over{2z}}\Bigg\rbrace\Bigg]
\equiv x(u)\quad ,\eqno(17b)
$$
after extracting the small-$z$ behavior.
The density is obtained by mass continuity~(2) and~(8):
\addtocounter{equation}{1}
\begin{eqnarray}
\rho (z) = \rho_c\exp\Bigl[-\int^{z}_0 {dz\ w(z)\over{[w(z)-z](3-z)}}\Bigr]
\quad .
\end{eqnarray}

\subsection{Central Density and Pressure Define Core Scale Factors}

To first order in $r^2$, hydrostatic equilibrium gives
\begin{equation}
P=P_c [1-(r/R_0)^2]\quad ,
\end{equation}
where $R_0^2\equiv 3P_c/2\pi G\rho_c^2$,
a central radial scale $R_0$ determined by the central pressure and density
and depending only implicitly on constitutive equations for equation 
of state, opacity, and energy generation.  We also
define $M_0\equiv (4\pi /3) R_0^3$, so that
$(P/\rho)_c = GM_0/2R_0, P_c=(2\pi G/3) R_0^2 \rho_c^2=(3G/8
\pi)(M^2_0/R^4_0)$.  Table~1 summarizes the BP98 central values of
$P,~\rho,$ and $P/\rho$ for the ZAMS and present SSMs, and the central
radial, mass and pressure scales derived therefrom.

Once $\rho_c$ and outer boundary condition are given,
$P_c$ is determined.  For given composition and constitutive
relations, stellar structure then depends on only one parameter,
the total mass $M$ (Vogt-Russell Theorem), 
assuming no critical instabilities such as the
Sch\" onberg-Chandrasekhar limit (Kippenhahn and Weigert 1990).
Integrating the equation of hydrostatic equilibrium,
\begin{eqnarray}
P(u) = P_c - \int^{m}_0{G m~dm \over{4\pi r^4}}
 = {G M^2\over{4\pi R^4}}\Big[\sigma_c - p(u)\Big]\quad ,
\end{eqnarray}
where $\sigma\equiv 4\pi R^4P(u)/GM^2$ is the (Schwarzschild)
dimensionless pressure defined by the total mass and radius, 
\begin{eqnarray}
\sigma_c\equiv 4\pi R^4P_c/GM^2\quad ,\quad
p(u) = \int^{3-u}_0 \Bigl[{q(z)\over{x(z)}}\Bigr]^2\cdot
{dz\over{[w(z)-z]}}\quad .
\end{eqnarray}

Assuming a zero-pressure boundary:
\begin{eqnarray}
\sigma_c=p(0) = \int^{3}_0 \Bigl[{q(z)\over{x(z)}}\Bigr]^2\cdot
{dz\over{[w(z)-z]}} =
{3\over{2}} \Big({\rho_c\over{\overline{\rho}}}\cdot{R_0\over{R}}\Big)^2\quad .
\end{eqnarray}
While $\rho_c/\overline{\rho}$ is a measure of {\em central}
concentration relative 
to the {\em whole} star, $\sqrt{\sigma_c}$ is a better measure of {\em core}
concentration.  This is how, for a given barytrope, the inner and outer
boundary conditions implicitly relate $P_c$ to $\rho_c$ or to $(R,M)$.
The $\sigma_c$ values for the BP98 ZAMS and present SSMs
are displayed at the bottom of Table~1.

For polytropes, $\rho_c/\overline{\rho}$ and $\sigma_c\equiv 4\pi W_n$
$\equiv 1/(n+1)[\theta^\prime_n(\xi_1)]^2$ are tabulated (Chandrasekhar 1939).
For $n$ = 0, 3, 4, 5, $\rho_c/\overline{\rho}$ and $\sqrt{\sigma_c}$
equal 1, 54.2, 622., $\infty$ and $\sqrt{3/2}$, 11.8, 55.8, $\infty$
respectively.  For the ZAMS (present) Sun, the
values $\sigma_c$ = 88.7 (260.) in Table~1, define equivalent values
$W_n$ = 7.06 (20.7) and polytropic indices $n$ = 2.796 (3.256), indices for
polytropes of the same core concentration $\sigma_c$.
For polytropes, the integral theorem~(22) reduces trivially to 1 =
$\theta^{n+1}_n(0) - \theta^{n+1}_n(\xi_1)$,
or $\theta_n(\xi_1)=0$, the
implicit definition of the dimensionless polytropic radius $\xi_1$.

\section{The Standard Solar Cores} 

Standard solar models use the best opacities and heavy element
diffusion to compute refined numerical results for solar neutrino fluxes and
helioseismology over the whole Sun.  Figures~2(a,b), 3(a,b) respectively show 
the $\rho$ and $P/\rho$ profiles for the BP98
ZAMS and present standard solar models (Pinsonneault 1998), which include 
helium and metal diffusion, both normalized to central
values. These figures also show our fits to the ZAMS (present) SSM cores
derived from our hyperbolic approximation
in Section~4 below.  The central values of pressure, density and 
$P/\rho$ for the ZAMS and present Suns are summarized in Table~1.  The values 
at the core radius are summarized in Table~2.
 
In the homogeneous ZAMS Sun, $1+\neff\equiv
d\ln P/d\ln (P/\rho)$ is identical to $1+N\equiv d\ln P/d\ln T\equiv
1/\nabla$ and varies from the almost adiabatic value 2.52 at $z$ = 0
to a maximum value 5.21 at $z$ = 0.32 $(r$ = 0.506 of ZAMS solar radius). 
Where the radiative zone meets the convective core, $1 + \neff$ shows
a discontinuity in slope.  The pressure-averaged value over the whole 
ZAMS Sun is $\neff\approx$ 3.

In the course of stellar evolution, the molecular weight gradient
$\nabla_{\mu}$ has increased from the original value $\nabla_{\mu}=0$
to the present value at the center, $\nabla_{\mu}=0.41$.
Meanwhile, the thermal gradient $\nabla$ has decreased, so that, at
present, $\Gamma\sim 1$ (Bludman \& Kennedy 1996). In the present
inner core, $P/\rho$ has a maximum ($v$ = 0 or $\Gamma=1$) at
$z_{\rm peak}$ = $0.36,~~m/M_\odot=0.034,~~r/R_{\odot} = 0.073$.  Inside
this zone, $\nabla_{\mu}>\nabla$ $(\Gamma < 1)$, 
and $w(u)$ has a slight negative curvature, seen in Figure~5(c).
$\nabla_\mu$ decreases radially from 0.41 at $z$ = 0 to 
zero at the edge of the core.
$\Gamma$ increases from $\Gamma_c$ = 0.86 to 1.31, then slowly declines
through most of the outer radiative zone.

For the {\em whole} ZAMS (present) Sun, 
the central concentration $\rho_c/\overline{\rho}$ = 35.3 (108)
would imply {\em global} effective polytropic indices 
$n\simeq$ 2.76 (3.35).  But the Eddington $n$ = 3 
Standard Model is a only crude approximation to the
whole present Sun.  The off-center $P/\rho$ peak shows that any polytropic 
approximation must fail badly in the present inner core, which shows 
{\em more} central concentration than does the whole Sun.  This inner core  
determines neutrino production and $g$-wave helioseismology.

\section{Analytic Approximations for Centrally-Concentrated Stellar Cores}

We now parametrize the BP98 SSMs by analytic formulas that incorporate the 
central boundary condition. Because they neglect the outer boundary condition
(9), these approximations are applicable only inside the core.
The mechanical structure of any star is represented by a $(u,w)$
curve asymptotic to $w = (5/3)(3 - u)=5z/3$ near the center $u=3,$
$w=0$ and curving upwards for $u\rightarrow 0$.  The straight-line behavior
is a good approximation only in the inner core $(u\gtrsim 2)$ of
centrally-concentrated cores, and needs to be corrected in the outer core.

\subsection{Hyperbolic Approximation}

Some of the upward curvature of $w(z)$ is captured by a simple
``hyperbolic'' approximation to the core of an extended polytrope
\begin{equation}
w(u) = (5/J)[1-(u/3)^J]\quad ,\quad w(z) = (5/J)[1-(1-z/3)^J]\quad ,
\end{equation}
where $J$ = $(9n-10)/7n$. 
At $u=0,~w=5/J$ is finite for any $J >$ 0, making the radius infinite.

Within the core, this one-parameter hyperbolic form is a closed approximation
to a polytrope.  The form~(23) is a Picard iterate (Boyce \&
DiPrima 1992) of equation~(7).  Integrating $w/n=-\xi\theta^\prime_n/\theta_n$,
equation~(23) gives
\begin{equation}
w(\xi )\approx (n\xi^2/3)/(1-\xi^2/6N)\quad ,\quad 
\theta_n(\xi )\approx [1-\xi^2/6N]^N\quad ,
\end{equation}
where $N\equiv 5/(5-3n)$.  This elementary
representation of Lane-Emden functions is exact for $n=0$ $(N=1,
J=-\infty )$ and for $n=5$ $(N= -1/2, J=1).$ For intermediate $n$, the
expansion of our approximation~(24) gives
\begin{equation}
\theta_n(\xi) \approx 1-\xi^2/6 + n\xi^4/120-n(6n-5)\xi^6/10800 +\cdots\quad ,
\end{equation}
which agrees with the exact Lane-Emden function
(Cox \& Giuli 1968)
\begin{equation}
\theta_n(\xi) = 1-\xi^2/6 + n\xi^4/120-n(8n-5)\xi^6/15120 +\cdots\quad ,
\end{equation}
through fourth order in $\xi$. The $\xi^6$ coefficient in the
hyperbolic approximation is $7(6n-5)/5(8n-5)$ times that in the exact
Lane-Emden function, or $0.985$ for $n=4$.  At least for polytropes
with $3 < n < 5$, the inner core $(z\leq 1)$ is well-approximated by the
hyperbolic approximation~(23).  The corresponding Taylor series
\begin{equation}
w(z) =(5z/3)[1+(5-n)z/21n + \cdots]
\end{equation}
converges only for small $z$.
The hyperbolic form is not intended to apply outside the core in
any case.

\subsection{Generalized Parametric Forms for Improved Pressure Structure}

If $\neff$ were constant, repeated Picard iteration of~(7) would improve
the hyperbolic approximation~(23), with better convergence for larger 
$z$ and better approximation to $\theta_n(\xi )$.
However, this refinement diverts us from considering a real star,
where $\neff$ is not constant.  This departure from homology
is especially important for describing helioseismology and neutrino production 
in the present solar core.  

In the non-polytropic case, equation~(7) becomes
\begin{eqnarray}
{d\ln (\Gamma w)\over{dz}}={2-z+w(\Gamma -1)\over{(3-z)(w-z)}}\quad .
\end{eqnarray}
The power of homology variables becomes evident when we construct
non-polytropic approximations to $w(z)$.
The most straightforward approach is to substitute an ansatz for
$\Gamma(z)$, integrate~(28) to obtain $w(z)$, and then deduce the 
desired structural profiles.  The simplest non-trivial ansatz,
a piecewise linear function for $\Gamma (z)$, suffices for the present Sun:
\begin{eqnarray}
\Gamma (z) = \left\{ \begin{array}{ll}
   \Gamma_c+(z/{z_{\rm peak}})(1-\Gamma_c)\quad ,& z<a\\
   \Gamma_c+(a/{z_{\rm peak}})(1-\Gamma_c)+b(z-a)\quad ,& z>a\\
                     \end{array} \right. \quad ,
\end{eqnarray}
where the parameters $\Gamma_c$, $z_{\rm peak}$, $a$, and $b$ obtained from
BP98 are given in Table~3.

Once equation (28) is solved for $w(z)$, from equation~(17b) and the definition
of $\Gamma$,
\begin{eqnarray}
r(z) & = & R_0\cdot [(5\Gamma_c/2)(z/3)]^{1/2}\cdot
\exp\Bigg[\int^{z}_{0}dz\Bigg\lbrace{1\over{(3-z)
[w(z)-z]}}-{1\over{2z}}\Bigg\rbrace\Bigg] \quad ,\nonumber \\
P(z) & = & P_c\cdot\exp\Big[-\int^{z}_0 {dz\ w(z)\ \Gamma (z)
\over{[w(z)-z](3-z)}}\Big]\quad .
\end{eqnarray}
All the non-trivial information and sensitivity of boundary conditions
is now contained in this pressure integral.

\section{Analytic Fits to the SSM Cores}

We now fit the preceding hyperbolic and parametric forms to the
entire core of BP98, which includes all of the energy and
neutrino producing parts of the Sun.  The BP98 models have finer zoning
near the center of the star, where, as we have shown, the mechanical 
structure is universally governed by mass continuity.  To correct for 
this over-representation of the central region and to make our fits 
sensitive to the non-trivial outer core
structure, we weight the BP98 data points inversely by their density in the
$(u,w)$ plane.  The local density is the number of model points per arc length
of $(u,w)$ curve.

Table~3 summarizes our best fits~(23,29) to the ZAMS and present Suns 
and their root mean variances (r.m.v.'s), the root of the sum of squared
deviations of fit and SSM model points, normalized by the degrees of freedom
in the fit.
Each hyperbolic fit can each be interpreted in terms of an 
equivalent stiffness $\GammaAV$ or
polytropic index $\nAV$ averaged over the core. Figures~4, 5 
show the BP98 data points and our hyperbolic 
(hyperbolic and parametric) fits in the ZAMS (present) Sun.

For the BP98 ZAMS SSM, Figures~4(a,b,c) show $u(m)$, $w(m)$, $u(w)$ 
together with our hyperbolic fit (dashed curve in Figure~4c) to the 94 data 
points in the 
radiative zone $u>1$. Because the innermost core of the ZAMS Sun has still 
not evolved far from its convective beginning $\Gamma\approx\Gamma_{\rm ad}$ =
5/3, the ZAMS core is approximated by a rather stiff $\nAV=1.8$ polytrope.
Figure 4(d) shows the variation of $\Gamma(z)$ over the ZAMS core.
Our hyperbolic approximation is sufficiently accurate:
using a general parametric form would lead to no improvement in the fit.

Figures~5(a,b,c) show the BP98 SSM present solar core, together with
our hyperbolic fit (dashed curve in Figure~5(c) to their 95 data points 
in the radiative zone $u>1$. Within its large uncertainty in $\nAV$,  
the hyperbolic fit (Table~3) is
consistent with the Eddington $n$ = 3 Standard Model. 
This {\em global} agreement with the Eddington model, however, misses the
crucial {\em local} feature of the present solar core.  
As seen in Figure~3b, the present Sun has evolved a broad off-center
maximum in $P/\rho$ at $r/R_{\odot}=0.073$, $m/M_{\odot}$= 0.034, which
cannot be described by any polytrope or other
one-parameter function.  Indeed, $\neff$ is infinite at this
maximum in $P/\rho$ and negative (positive) inside (outside) this
zone, explaining the large uncertainty in $\nAV$ obtained by the simple
hyperbolic fit.
Although $\neff$ is singular, $\Gamma$ varies smoothly through unity
at the peak in $P/\rho$ at $z_{\rm peak}$ = 0.36, 
as shown in Figure~5(d).

The parametric best fit in Table~3 is
six times better than the hyperbolic best fit.
The resulting improved profiles of $w$, $\rho$, and $P$ are shown 
by the solid curves of Figures~3 and~5. 
The improvement in the double logarithmic derivative $w(u),$ although 
it appears small to the eye, is important: with it, we get an excellent fit to 
$P/\rho$ in the present solar core. 

\section{Analytic Fit to Helioseismological Observations}

We now fit to helioseismic data from 
the present Sun and the inferred adiabatic squared sound speed $c^2_{\rm ad} 
=\Gamma_{\rm ad} \cdot (P/\rho)$, where $\Gamma_{\rm ad}\equiv 
(\partial\ln P/\partial\ln\rho)_{\rm ad}$.
Hydrostatic equilibrium~(3) gives the the {\rm slope} of the sound speed:
\be
{dc^2_{\rm ad}\over{dr}} = -g\Gamma_{\rm ad}\Big(1-\Gamma^{-1}\Big) +
{P\over\rho}\cdot{d\Gamma_{\rm ad}\over{dr}}\quad .
\ee
In the inner core of the present Sun, the local $g$ is well-determined by
mass continuity, but $(1-\Gamma^{-1})$ = $1/(1+\neff )$ varies greatly,
because of the evolved molecular weight gradient $\nabla_\mu$.

The $p$-mode sound speed (Figure 6, Basu {et al.} 1997, and 
Christensen-Dalsgaard 1997) has been inferred from comparison of solar models 
(Christensen-Dalsgaard {et al.} 1985) and helioseismic observations 
now reaching virtually to the solar center (Christensen-Dalsgaard {et
al.} 1996; Gough {et al.} 1996; Kosovichev {et al.} 1997; Basu {et al.} 1997;
GONG Collaboration 1999; SOI Collaboration 1999).
Assuming only spherical symmetry, mass continuity, pressure equilibrium
and full ionization in the core $(\Gamma_{\rm ad}$ = 5/3), 
$dc^2_{\rm ad}(0)/dr$ must vanish, and we fit the 22 points $r <$
$(0.32)R_\odot$ in Figure~6 to a polynomial in $r^2$ to obtain the central
sound speed.  Our best fit is ninth
order in $r^2$ and gives
\begin{eqnarray}
c_{\rm ad}(0) = (5.07\pm 0.09)\times 10^7\ {\rm cm\ sec}^{-1}\quad ,
\end{eqnarray}
in excellent agreement with the BP98 present Sun SSM value of $5.053\times 
10^7$ cm sec$^{-1}$ (negligible error).  $c^2_{\rm ad}(0)$ fixes $\sigma_c
(\overline{\rho}/\rho_c)$ = $R_\odot c^2_{\rm ad}(0)/GM_\odot\Gamma_{\rm ad}$ =
$0.81\pm 0.03$.

\section{Conclusions}

The BP98 ZAMS and present SSMs are well described by the analytic and
semi-analytic formulas~(23,24,28--30,18).  Because it is still partly 
convective, the core of the ZAMS Sun (Figures~2, 4) is globally
approximated, with root mean variance 0.1, by a stiff polytrope of
$\GammaAV = 1.6\pm 0.1$, $\nAV = 1.8\pm 0.2$.

The core of the present Sun (Figures~3, 5) is only crudely polytropic
with index $\nAV = 3.4\pm 1.0$ and root mean variance 0.1 (Figure~5).  
The inhomogeneous inner core requires a
variable $1-\Gamma^{-1} = \nabla -\nabla_\mu$.  A parametric form
that generalizes the polytrope to a linearly varying $\Gamma$ 
accurately fits the present solar core.

Helioseismological data can be analyzed in the same phenomenological
approach, ultimately
extensible to a fully model-independent framework.  Present data do not
allow an {\em independent} reconstruction of inner core structure,
but future measurements of $g$ modes and refinement of $p$-mode data should 
allow further iteration of full numerical models or refinement of 
our parametric approach~(28--31) with varying 
$\Gamma$.  Asteroseismology can be similarly analyzed.

\acknowledgments

We thank Marc Pinsonneault of Ohio State University for the BP98 SSM tables,
and John Bahcall of the Institute for Advanced Study, Princeton,
and J{\o}rgen Christensen-Dalsgaard of the Institute of Physics and
Astronomy, Aarhus University, Denmark, for helioseismology
data. This work was supported by the U.S. National Science Foundation under
grant PHY89-04035 (U. California, Santa Barbara), 
the U.S. Department of Energy
under contracts DE-FG05-86-ER40272 (Florida) and DE-AC02-76-ERO-3071
(Penn), and by the Institute for Fundamental Theory (U.~Florida).

\vskip 0.5in
\newpage

\begin{table}
\centering
\begin{tabular}{||lr|c|c||}  \hline
           &           &  {\em ZAMS Sun}  & {\em Present Sun} \\  \hline 
$P_c$ & $(10^{17}$ dyn cm$^{-2})$  &  1.37  &  2.33         \\
$\rho_c$ & (gm cm$^{-3})$  &  74.7  &  152.  \\
$P_c/\rho_c$ & $(10^{15}$ cm$^2$ sec$^{-2})$  &  1.83  &  1.53  \\
$R_0/R$ & &  0.218  &  0.122  \\
$M_0/M$ & &  0.366  &  0.196  \\  \hline
$GM^2/4\pi R^4$ & $(10^{15}$ dyn cm$^{-2})$  &  1.54  & 0.895  \\ 
$\sigma_c$ &  &  88.7  &  260.  \\
$\rho_c/\overline{\rho}$ & &  35.3  &  108.  \\  
$n$ & & 2.796 & 3.256 \\ \hline
\end{tabular}
\bigskip
\caption{Central pressure, density, pressure scale height, and other
features of the BP98 ZAMS and present SSMs.
The global quantities below the horizontal line describe ZAMS and present
Suns of total mass $M$ = $M_\odot$ = $1.989\times 10^{33}$ gm and
radii $R$ = $(0.873)R_\odot$, $R_\odot$ = $6.96\times 10^{10}$ cm
respectively.  $\rho_c/\overline{\rho}$ and $\sigma_c$ are measures of
central and
core concentration for the entire Suns.  $n$ 
is the polytropic index of a complete polytrope showing same value of 
$\sigma_c$.}\end{table}

\begin{table}
\centering
\begin{tabular}{||l|c|c||}  \hline
          &  {\em ZAMS Sun}  &  {\em Present Sun}  \\  \hline
$R_0/R$   &  0.218  &  0.122  \\
$r_g/R_0$ &  1.86   &  2.67   \\
$r_g/R$   &  0.404  &  0.326  \\
$m_g/M$   &  0.711  &  0.666  \\
$L_g/L$   &  1.00   &  1.00   \\  
$w_g$     &  4.05   &  3.67   \\  \hline          
\end{tabular}
\bigskip
\caption{Entire core mass $m_g$, radius $r_g$, luminosity $L_g$ and homology
invariant $w_g$ of BP98 ZAMS
and present SSMs.  Total mass, radii, and luminosities of the ZAMS and 
present Suns are $M$ = $M_\odot$ = $1.989\times 10^{33}$ gm; $R$ = 
$(0.873)R_\odot$, $R_\odot$ = $6.96\times 10^{10}$ cm; and $L$ = 
$(0.679)L_\odot$, $L_\odot$ = $3.84\times 10^{33}$ erg sec$^{-1}$
respectively.  Entire core is defined by $u_g$ = 1.}
\end{table}

\begin{table}
\centering
\begin{tabular}{||c|c|c|c||}  \hline
{\em Type of Fit} & &  {\em ZAMS Sun}  &  {\em Present Sun}  \\ \hline
  & $J$    &  $0.5\pm 0.1$           &  $0.9\pm 0.1$      \\ \cline{2-4}
Hyperbolic & $\nAV$ & $1.8\pm 0.2$   &  $3.4\pm 1.0$      \\ \cline{2-4}
  & $\GammaAV$ & $1.6\pm 0.1$  &  $1.3\pm 0.1$            \\ \cline{2-4}
  & r.m.v. &  0.097                  &  0.10              \\ \hline
Improved & $\Gamma_c$ & 1.66 &  $0.86$        \\ \cline{2-4}
General & $z_{\rm peak}$ & --- & $0.33$\\ \cline{2-4}
Para- & $a$ & --- & $1.09\pm 0.01$ \\ \cline{2-4}
metric & $b$ & --- & $-0.07\pm 0.02$ \\ \cline{2-4}
       & r.m.v. & --- & 0.016                   \\ \hline
\end{tabular}
\bigskip
\caption{One-parameter hyperbolic and four-parameter generalized parametric
fits to BP98 ZAMS and present SSMs.  
$\nAV$ and $\GammaAV$ = $1 + 1/\nAV$ are index and
stiffness of polytropes having the same {\em average} core properties.
Below the line,
the four parameters describing variation 
of $\Gamma (z)$ give a much improved fit for present solar core.}
\end{table}

\newpage
\centerline{\bf FIGURES}

\figcaption{
The characteristic $w(u)$ curves of the mass and density 
logarithmic derivatives, for complete polytropes $n$ = 0, 1, 2, 3, 4, 5,
$\infty$.  All stars satisfy the central boundary condition $w \rightarrow 
(5/3)(3-u)$.
Centrally-concentrated stars $(n > 3)$ show this common structure
inside an inner core defined by $2\le u\le 3$ (up to maximum 
gravitational acceleration). The total mass and radius are finite for each 
polytrope $n<5$.  The $n=5$ polytrope has infinite
radius for every finite mass.\label{FIG1}}

\figcaption{
The mechanical structure of the BP98 ZAMS Sun: (a) density with central
value $\rho_c=74.7$ g cm$^{-3}$ and (b) pressure-density ratio with central
value
$(P/\rho)_c= 1.83\times 10^{15}$ cm$^2$ sec$^{-2}$.  Dashed curves: same
properties for our hyperbolic best fit. 
Vertical arrows: outer core radius defined by $u=1$. \label{FIG2}}
 
\figcaption{
The mechanical structure of the BP98 present Sun: (a) density with central
value $\rho_c=152.$ g cm$^{-3}$ and (b) pressure-density ratio with central
value
$(P/\rho)_c= 1.53\times 10^{15}$ cm$^2$ sec$^{-2}$.  Dashed curves: same
properties for our hyperbolic best fit.
Solid curves: same properties for our parametric best fit.
Vertical arrows: outer core radius.\label{FIG3}}

\figcaption{
The BP98 ZAMS Sun in terms of homologous variables $(u,w)$: (a) $u$ as function
of mass fraction $m/M_\odot$; (b) $w$ as function of mass fraction
$m/M_\odot$; (c) parametric plot of $w(u);$ (d) variable $\Gamma (z)$.
Our hyperbolic best fit (dashed curves) is an adequate fit to the ZAMS core.
\label{FIG4}}

\figcaption{
The BP98 present Sun in terms of homologous variables $(u,w)$: (a) $u$ as
function of mass fraction $m/M_\odot$; (b) $w$ as function of mass fraction
$m/M_\odot$; (c) parametric plot of $w(u);$ (d) variable $\Gamma (z)$.
Our hyperbolic best fit (dashed curves) is unacceptable.
Our parametric best fit (solid curves) is much better than the 
hyperbolic fit to the present solar core.\label{FIG5}}
  
\figcaption{
The $p$-mode adiabatic sound speed in core of Sun inferred by linearized 
perturbative inversion from helioseismic observations and a reference 
SSM.  Uncertainties too small to show.  Solid curve: our parametric 
best fit.
(Data courtesy of J.~Christensen-Dalsgaard.)\label{FIG6}}

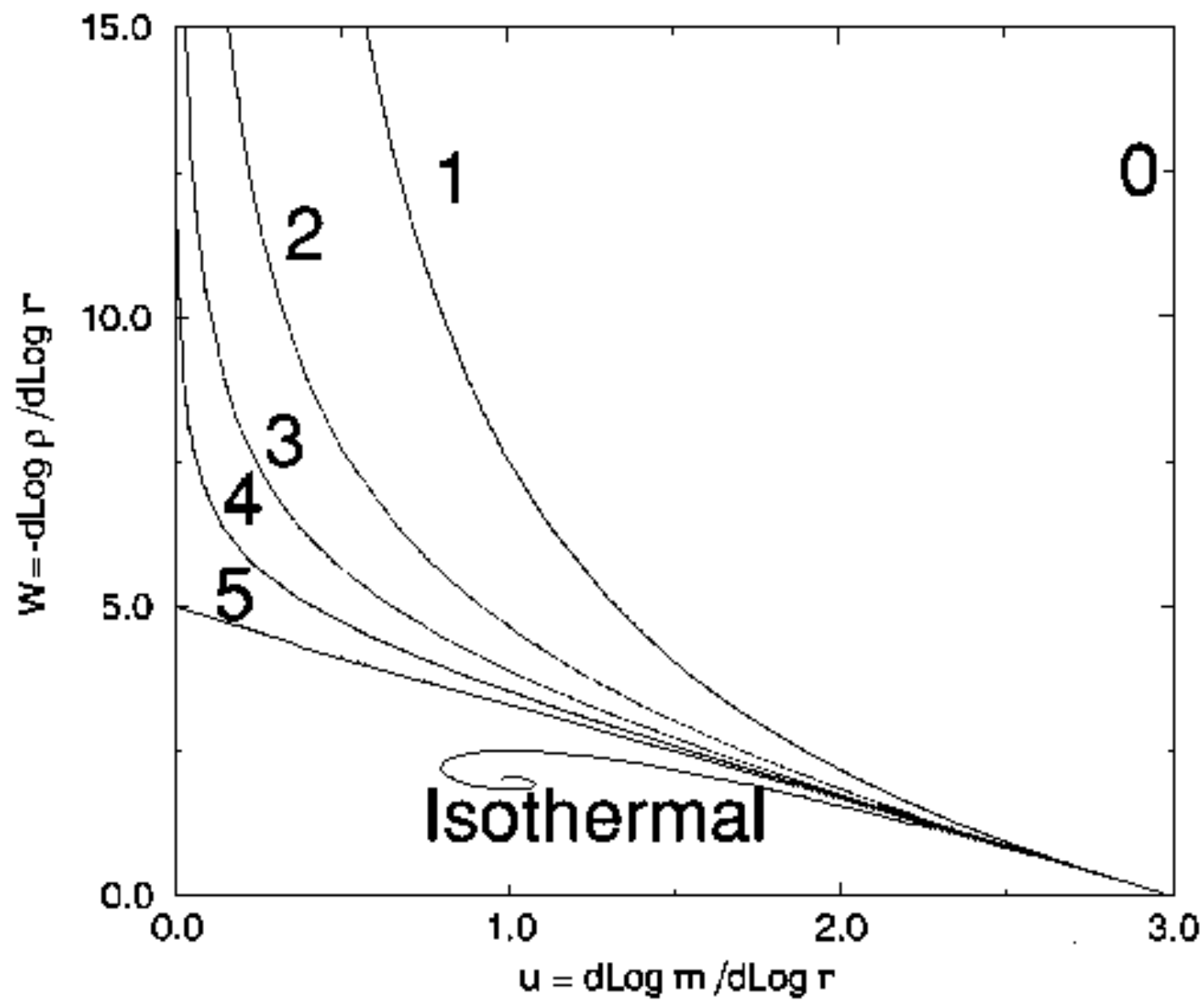
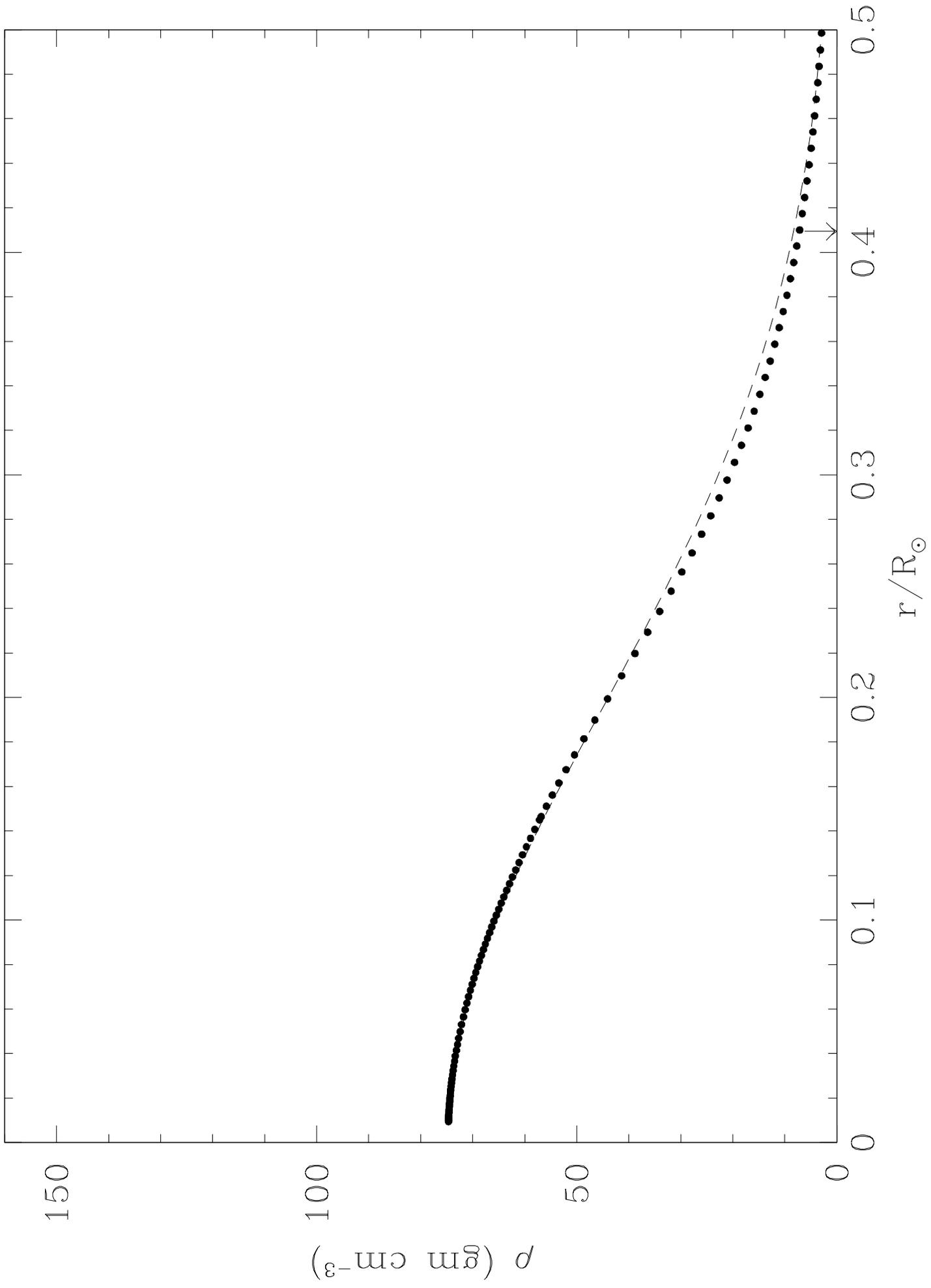
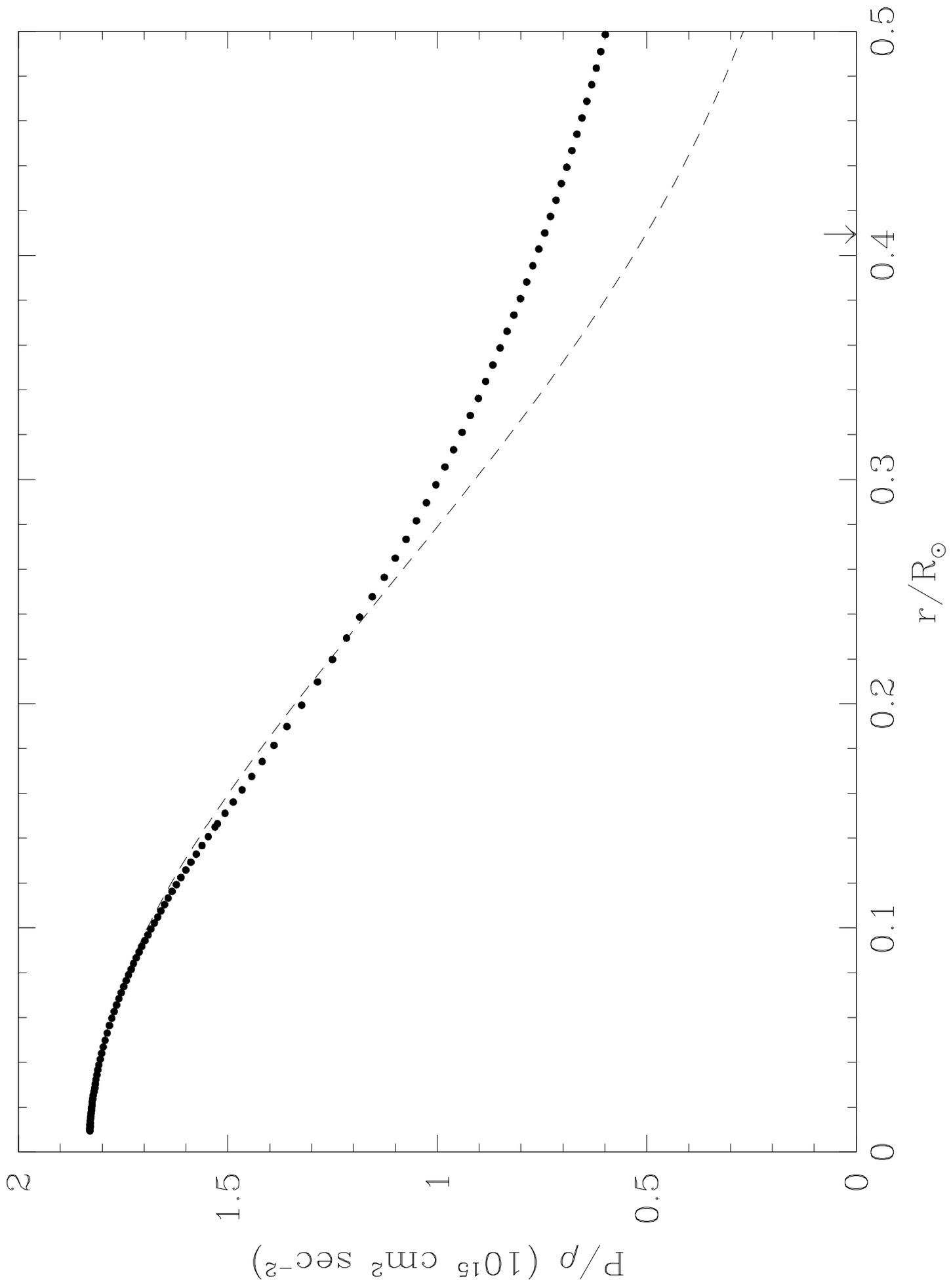
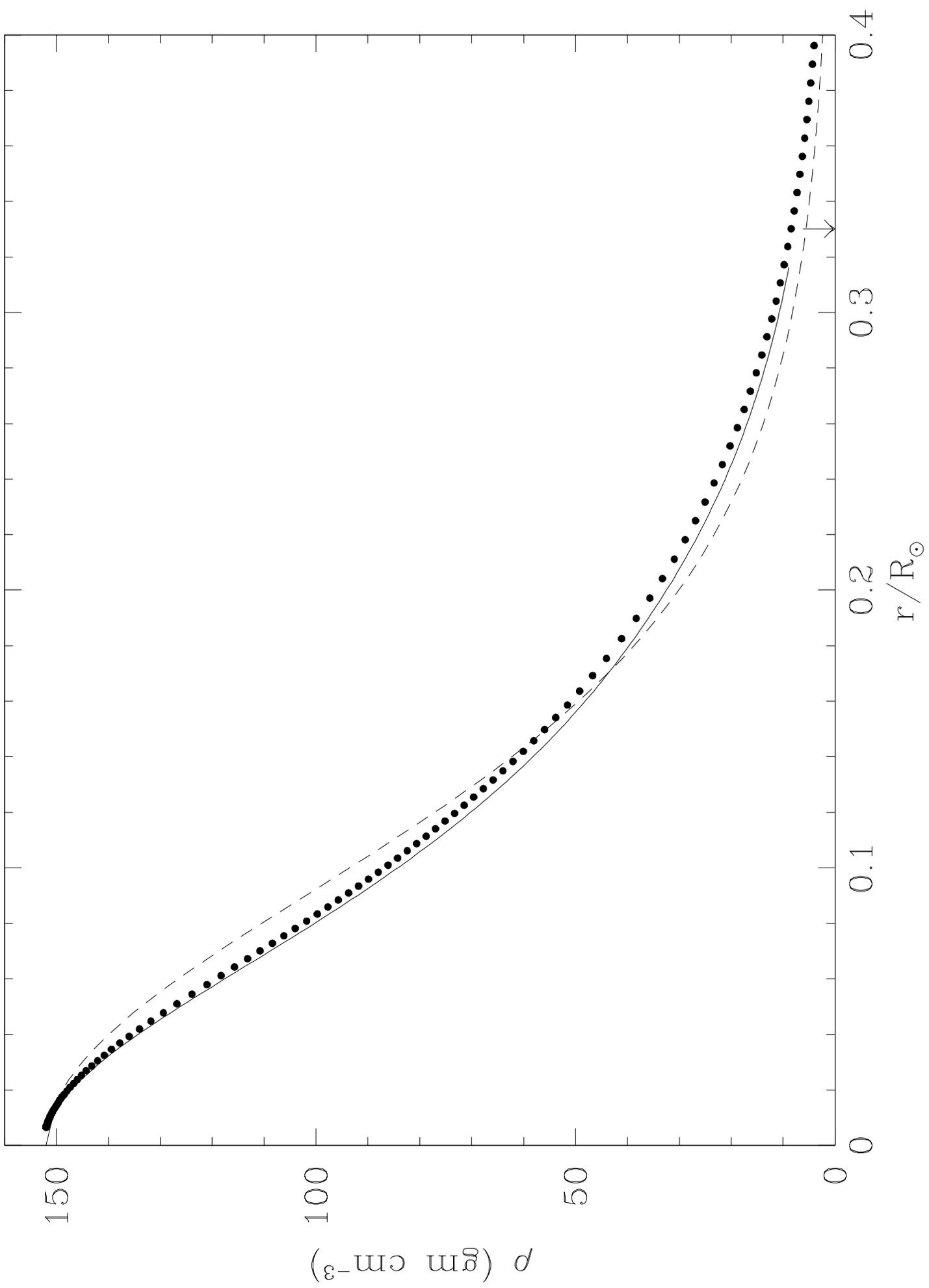
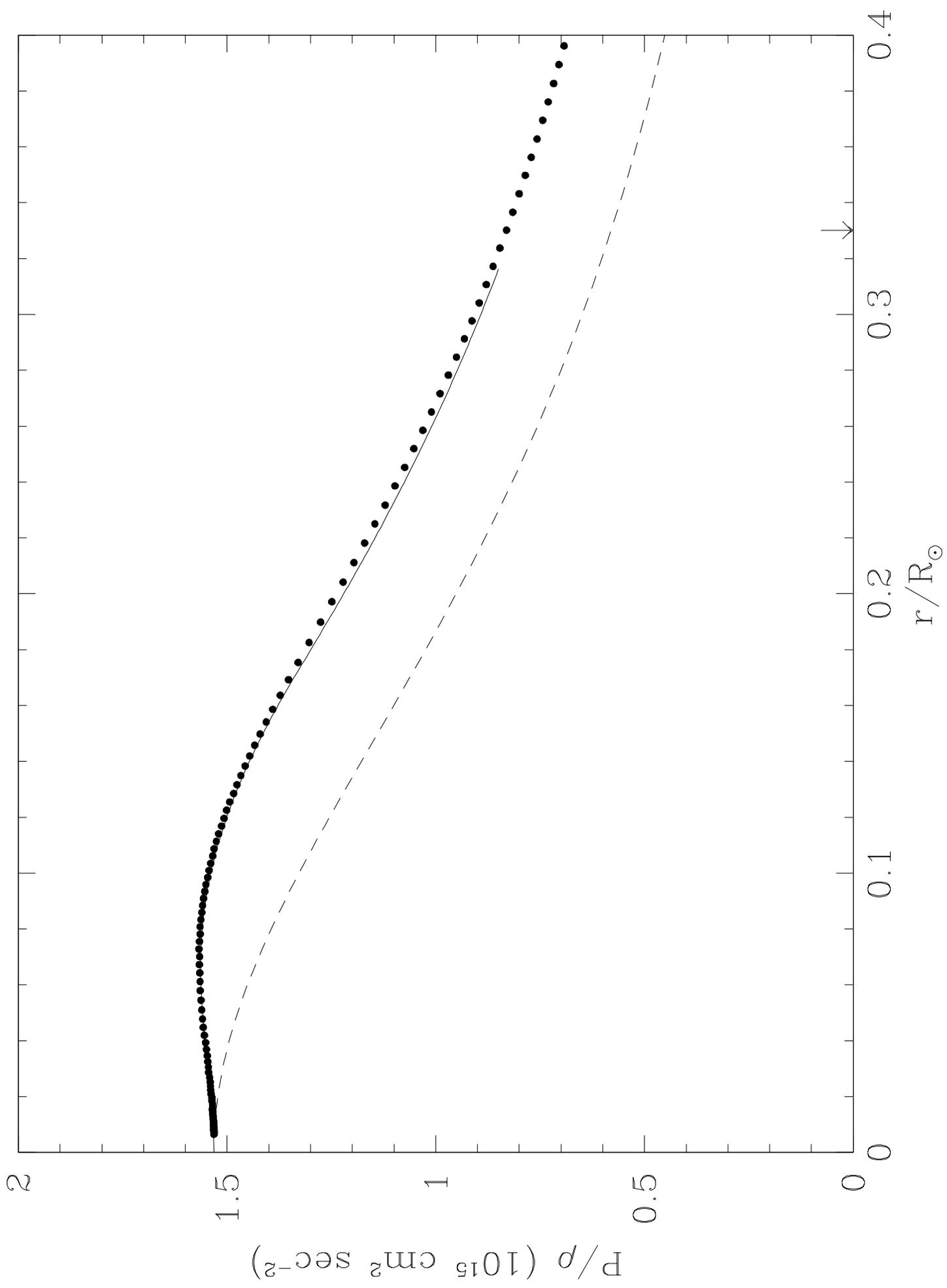
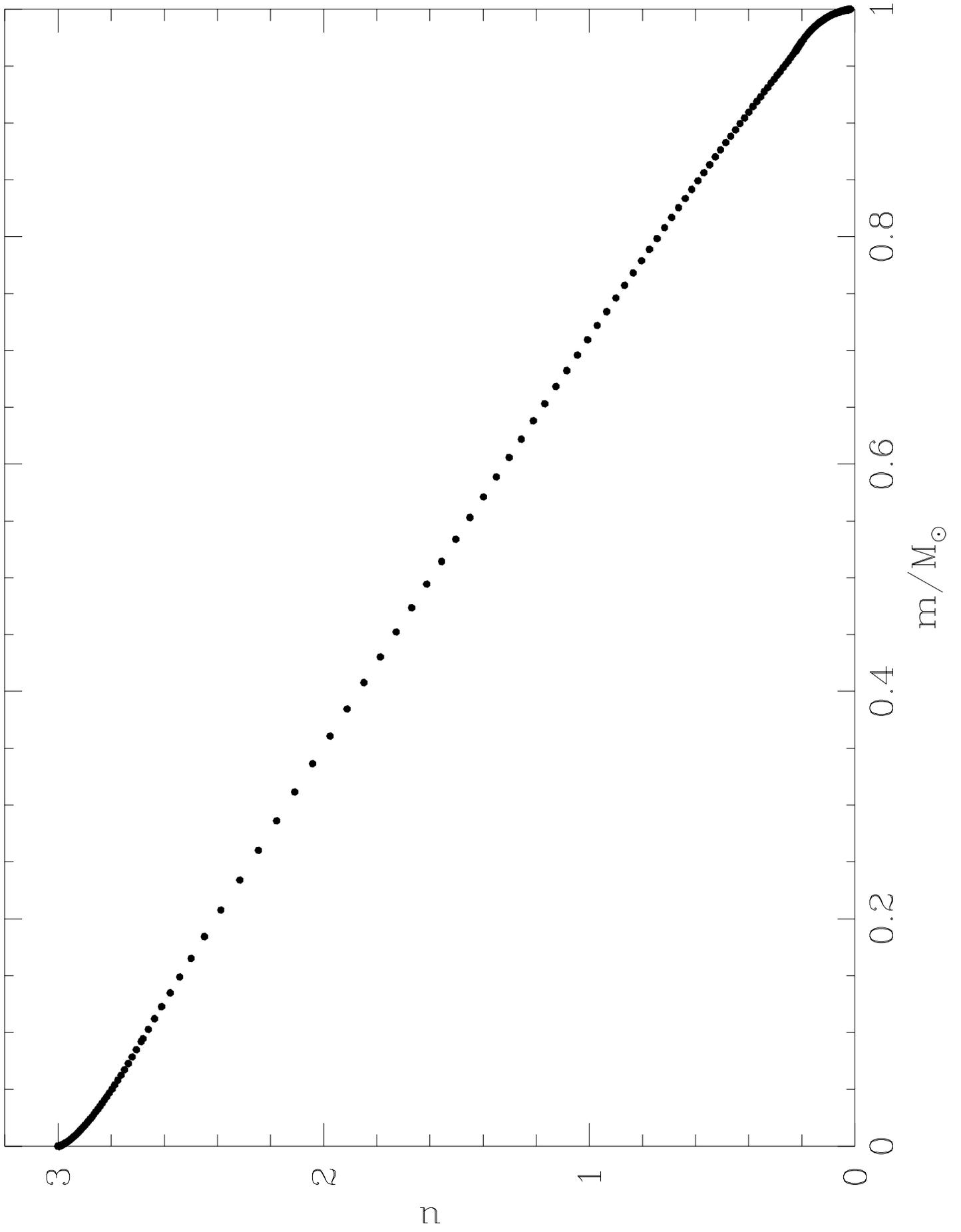
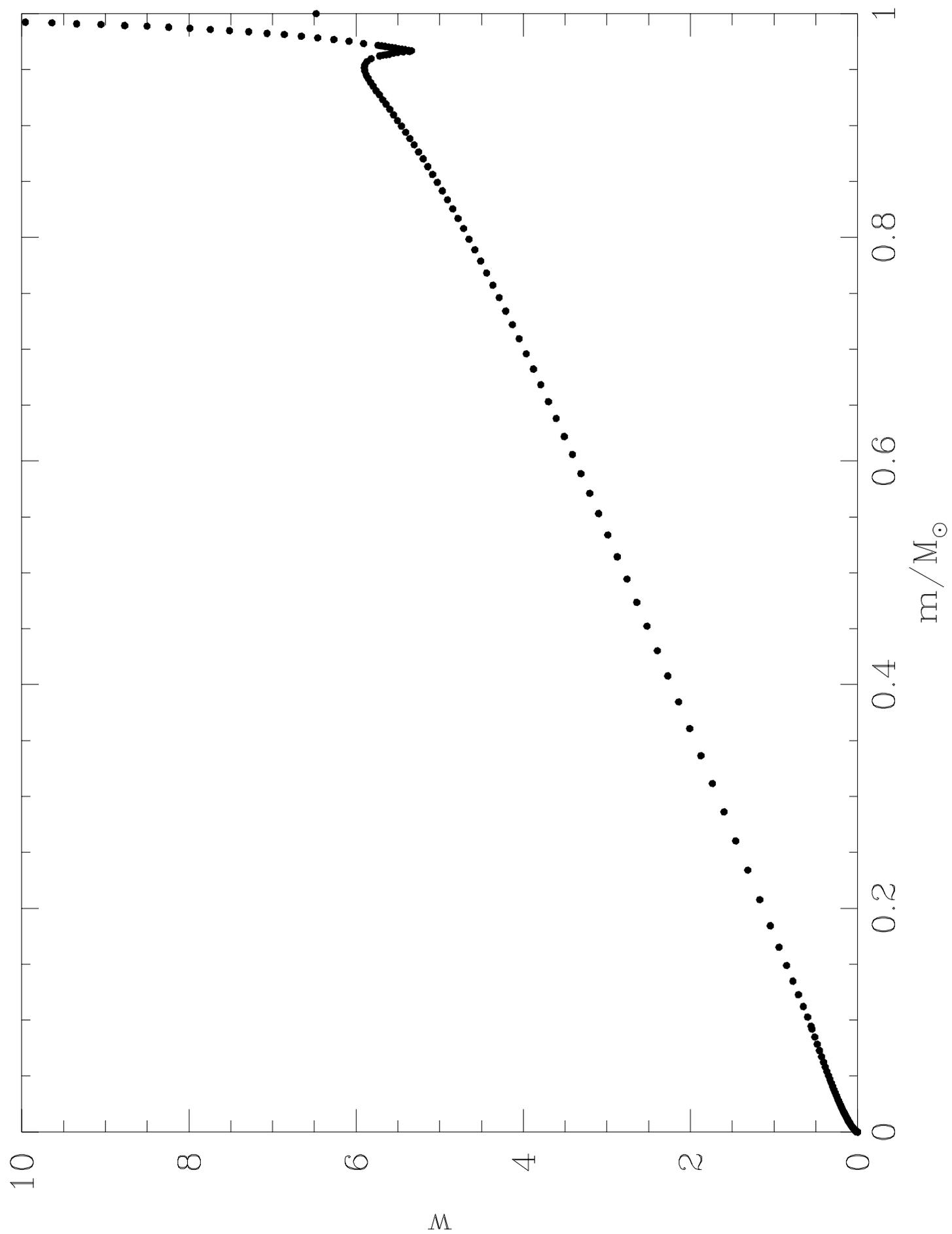
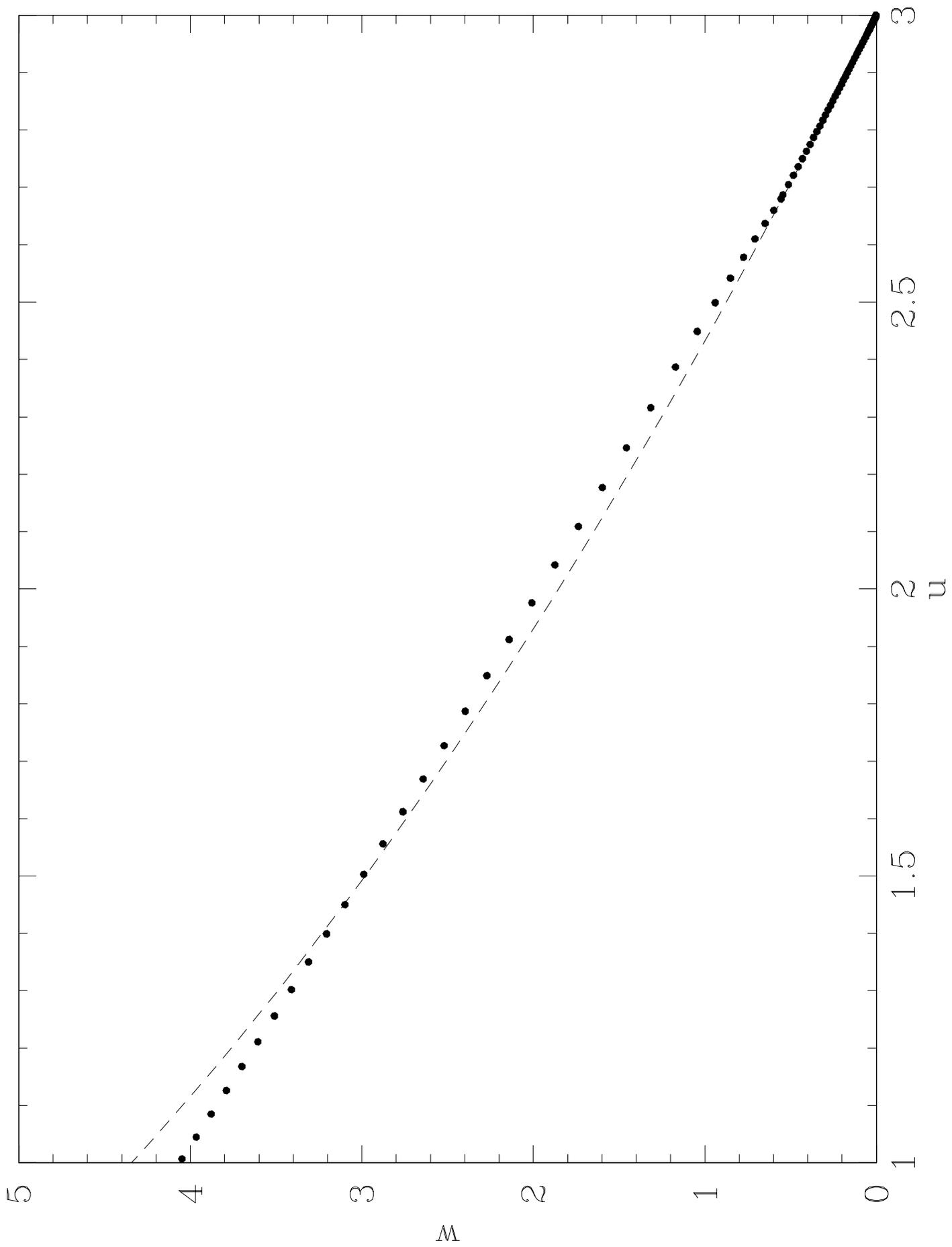
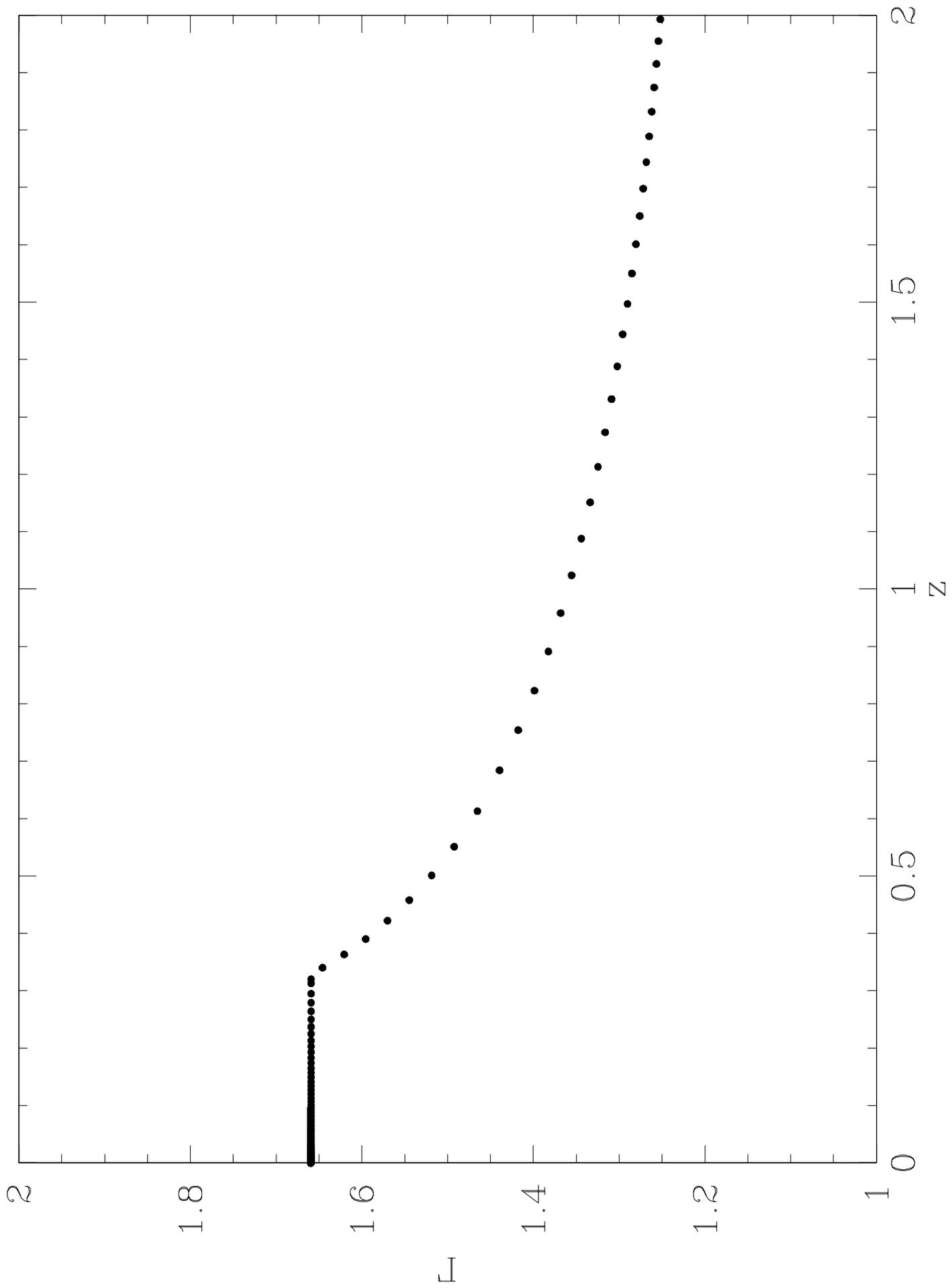
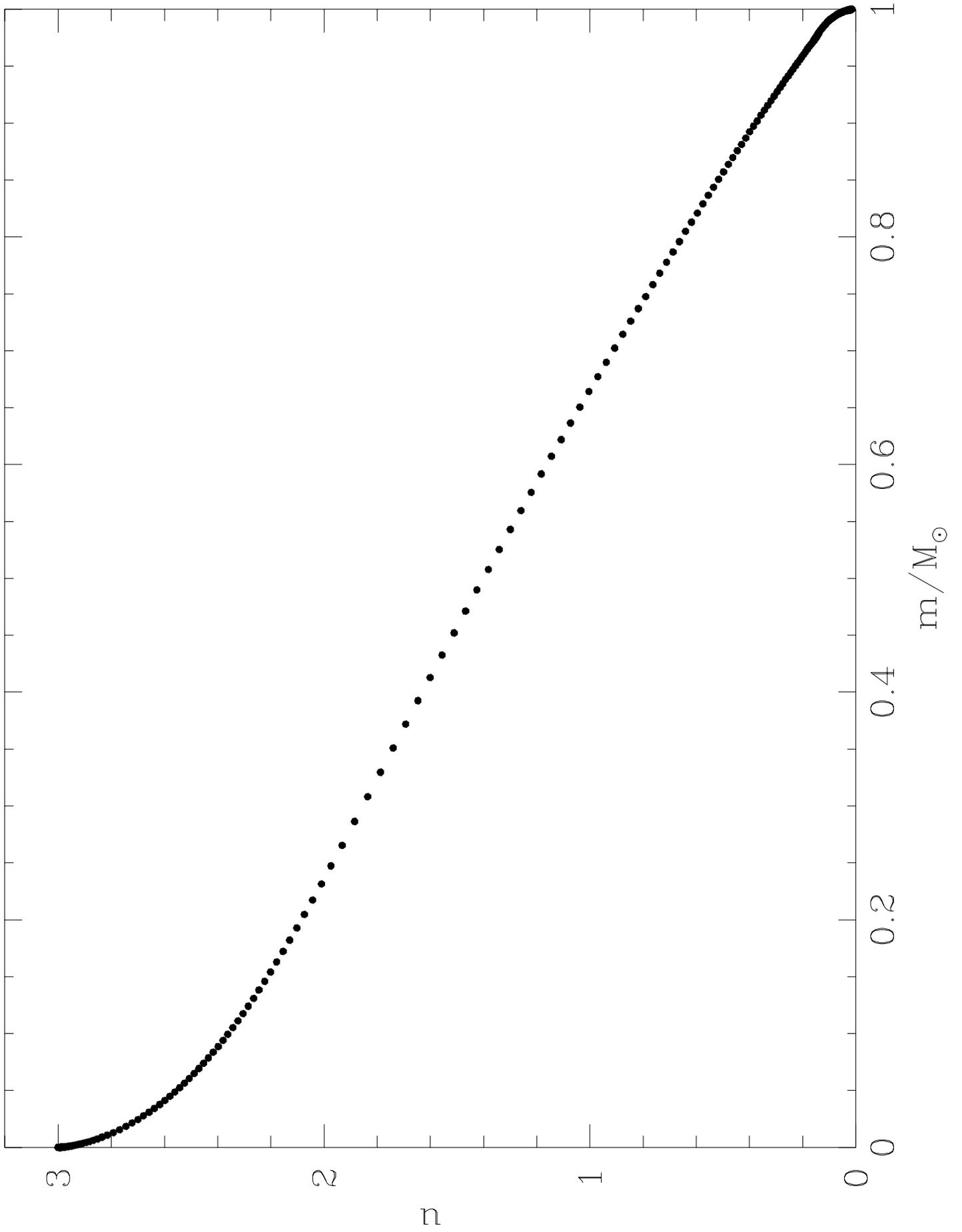
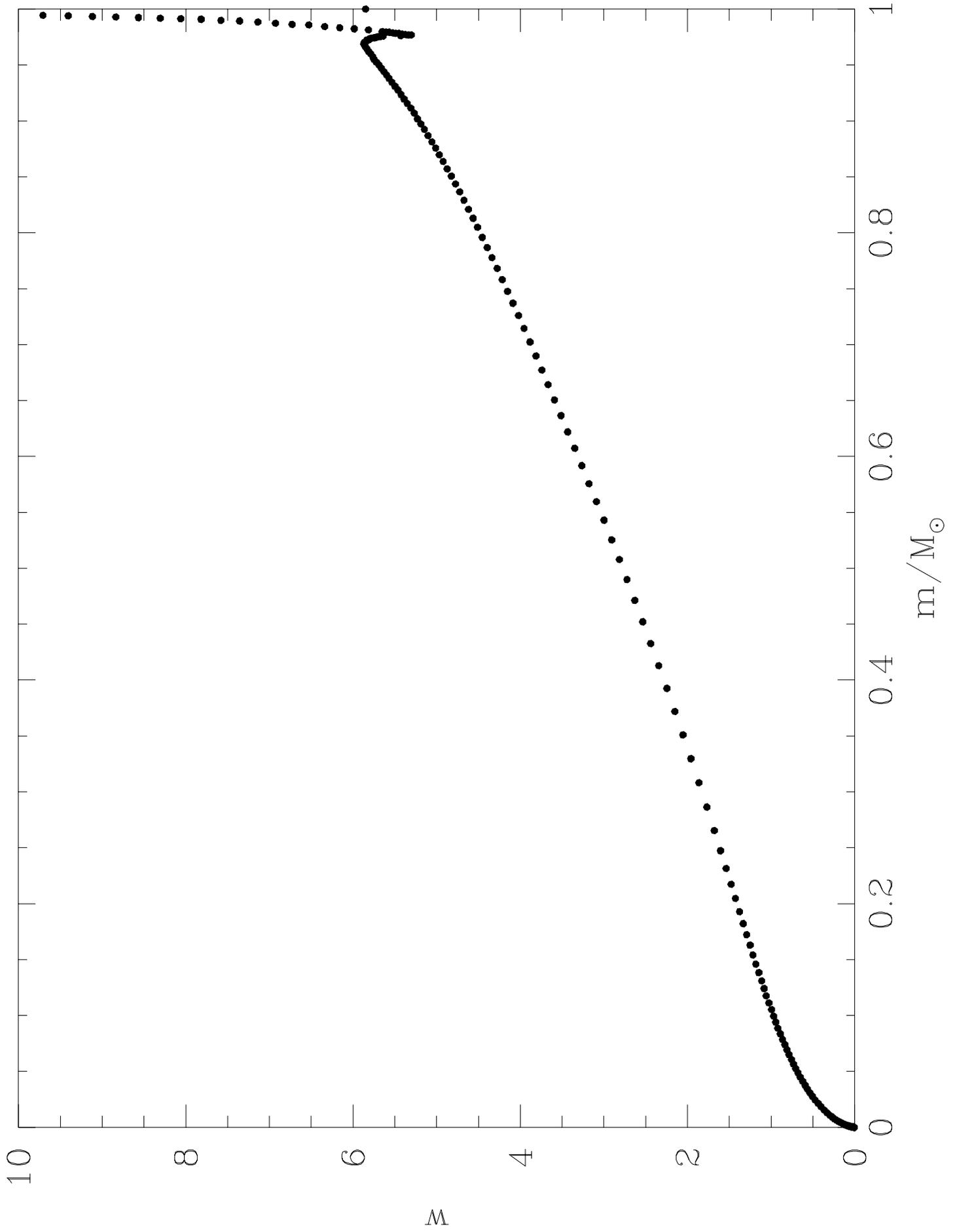
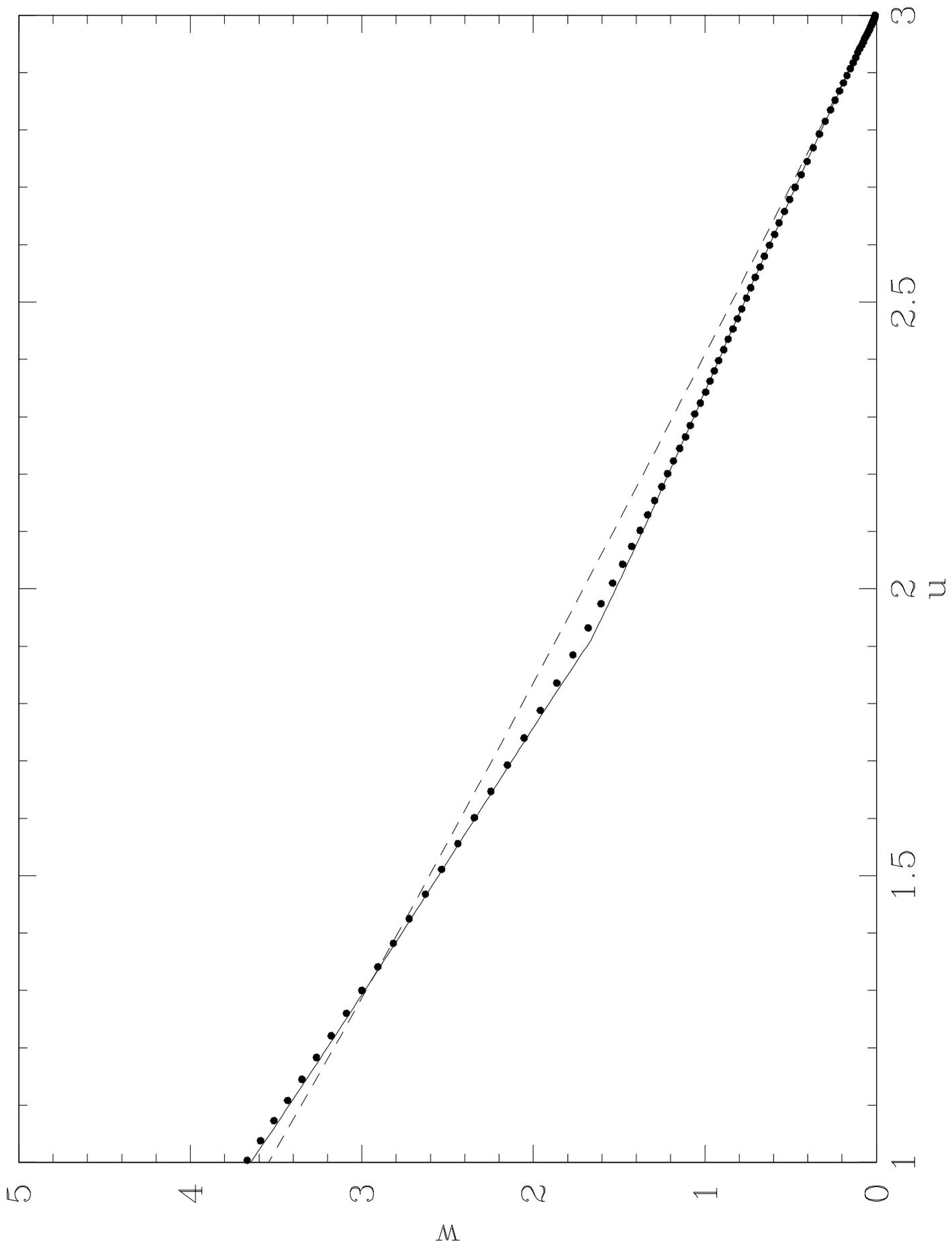
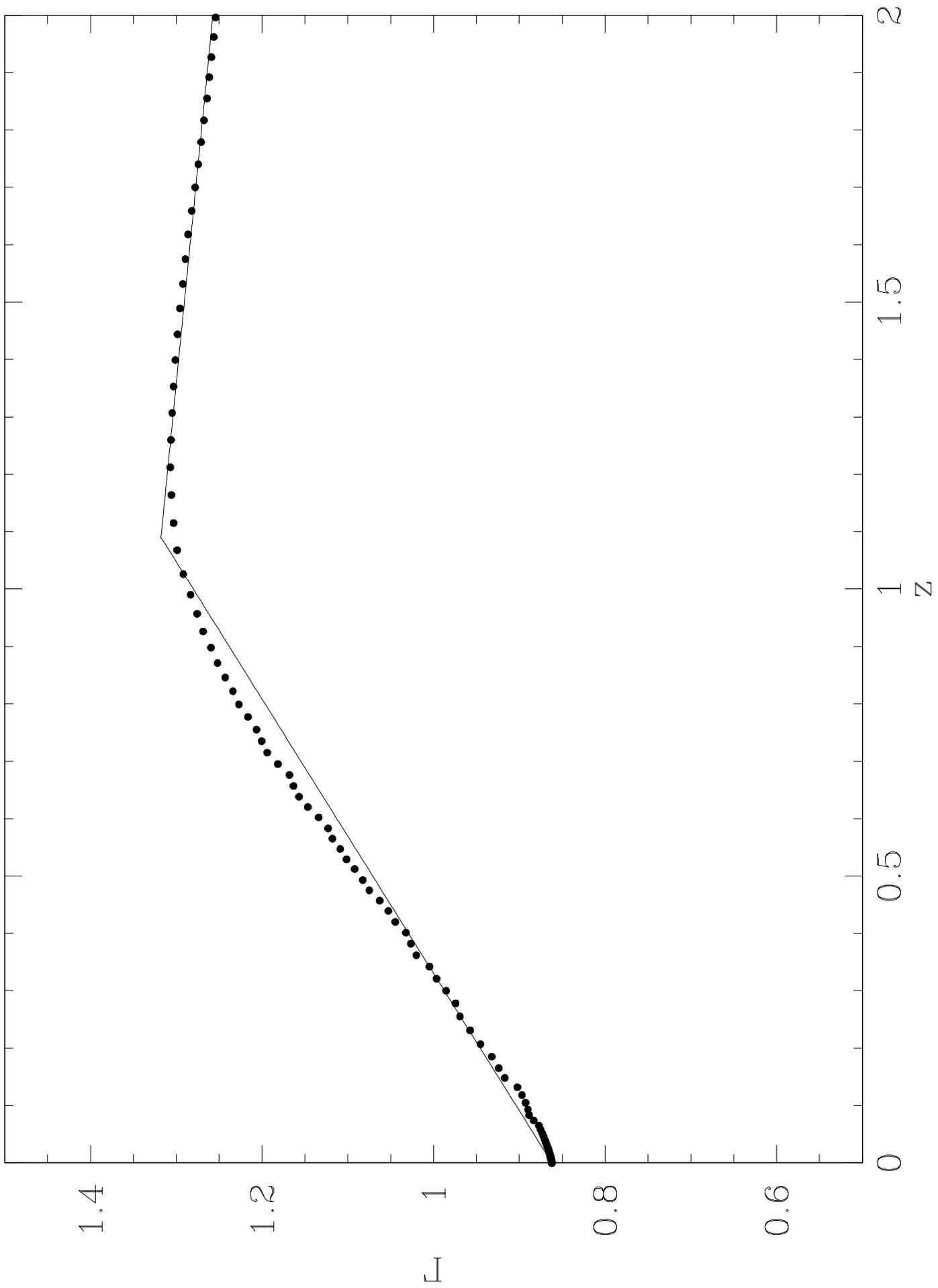
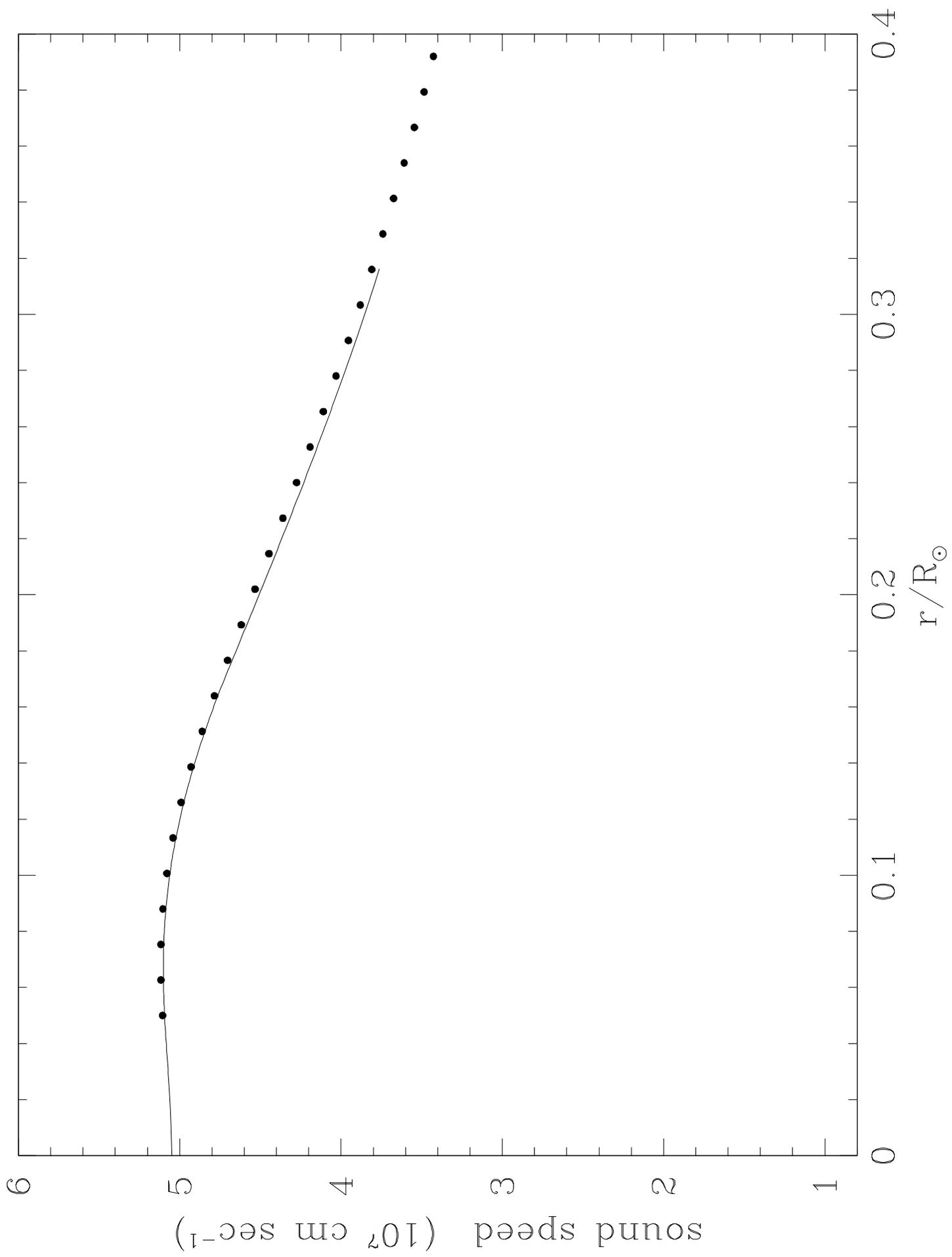
\end{document}